\documentclass[sigconf]{acmart}
\usepackage[utf8x]{inputenc}

\usepackage{graphicx}
\usepackage{tabularx}
\usepackage{multirow}

\usepackage{comment}

\pdfoutput=1

\def\BibTeX{{\rm B\kern-.05em{\sc i\kern-.025em b}\kern-.08emT\kern-.1667em\lower.7ex\hbox{E}\kern-.125emX}}
    
\setcopyright{none}

\begin{document}

%
% The "title" command has an optional parameter, allowing the author to define a "short title" to be used in page headers.
\title{Wikidata from a Research Perspective - A Systematic Mapping Study of Wikidata}

%
% The "author" command and its associated commands are used to define the authors and their affiliations.
% Of note is the shared affiliation of the first two authors, and the "authornote" and "authornotemark" commands
% used to denote shared contribution to the research.
\author{Mariam Farda-Sarbas}
\email{mariam.farda.sarbas@fu-berlin.de}
\author{Claudia Müller-Birn}
\email{clmb@inf.fu-berlin.de}
\affiliation{%
  \institution{Human Centered Computing | Freie Universität Berlin}
  \streetaddress{Königin-Luise-Str. 24-26}
  %\city{Berlin}
  %\country{Germany}
  \postcode{14195}
}
%
% By default, the full list of authors will be used in the page headers. Often, this list is too long, and will overlap
% other information printed in the page headers. This command allows the author to define a more concise list
% of authors' names for this purpose.
\renewcommand{\shortauthors}{Farda-Sarbas and Müller-Birn}

%
% The abstract is a short summary of the work to be presented in the article.
\begin{abstract}
Wikidata is one of the most edited knowledge bases which contains structured data. It serves as the data source for many projects in the Wikimedia sphere and beyond. Since its inception in October 2012, it has been increasingly growing in term of both its community and its content. This growth is reflected by an expanding number of research focusing on Wikidata. 
Our study aims to provide a general overview of the research performed on Wikidata through a systematic mapping study in order to identify the current topical coverage of existing research as well as the white spots which need further investigation. 
In this study, 67 peer-reviewed research from journals and conference proceedings were selected, and classified into meaningful categories. We describe this data set descriptively by showing the publication frequency, the publication venue and the origin of the authors and reveal current research focuses. These especially include aspects concerning data quality, including questions related to language coverage and data integrity. These results indicate a number of future research directions, such as, multilingualism and overcoming language gaps, the impact of plurality on the quality of Wikidata's data, Wikidata's potential in various disciplines, and usability of user interface.

\end{abstract}



%
% Keywords. The author(s) should pick words that accurately describe the work being
% presented. Separate the keywords with commas.

%
% This command processes the author and affiliation and title information and builds
% the first part of the formatted document.
\maketitle
\section{Introduction}
\label{sec:intro}

Wikidata, the sister project of Wikipedia, is a collaborative knowledge base (KB)\footnote{A knowledge base is a centralized repository of data, which stores data in any form, such as in a tabular or graph format.}, which is openly accessible and contains human-readable, as well as, machine-readable data. Wikidata was launched in October 2012 and since then, it has been one of the most often edited knowledge bases with around 20,000 active users\footnote{https://www.wikidata.org/wiki/Wikidata:Statistics}. The main goal behind Wikidata's development is to provide structured data for Wikimedia projects to overcome the data inconsistencies of Wikipedia's language versions. Wikidata is designed in a way that anyone can edit, browse, consume, and reuse the data in a fully multilingual form~\cite{vrandecic_rise_2013}. 

Wikidata’s content is also stored as a knowledge graph (KG)\footnote{We refer to knowledge graph when we mean a knowledge base which stores the data in graph format.}. Thus, the data are provided in a structured form in the RDF\footnote{RDF is the abbreviation for Resource Description Framework.} format and can be accessed using SPARQL\footnote{SPARQL is the abbrviation for Simple Protocol and RDF Query Language.}.  Wikidata is designed a) to be open for anyone to edit (with or without Wikidata account), b) to allow conflicting ideas to coexist, c) to provide data in a language independent form, d) to be controlled by the contributing community, e) to provide data with references, and f) to continuously evolve to address the emerging needs of the users and contributors~\cite{vrandecic_wikidata:_2014}. 

As a multilingual knowledge base, Wikidata can provide data in any context and language as long as it is available. Thus, the provided data is already being used by other projects such as, WDAqua-core\footnote{A question answering service for RDF knowledge bases~\cite{diefenbach_wdaqua-core1:_2018}.}, WikiGenomes\footnote{A community created for consuming and curation gene annotation data~\cite{e._putman_wikigenomes:_2017}.} and Open Street Map\footnote{A peer production project that creates an editable global map~\cite{leyh_interlinking_2017}.} and the increasing number of contributors and contributions show the rising interest in Wikidata. 

At the same time, the research community interest on Wikidata has accumulated recently, and this is an indication of its growing popularity. Numerous studies have explored Wikidata from various angles, such as its internal structure, including both, data and community, from a data perspective by looking at its completeness and coverage, from a engineering perspective by looking at the needed tools, and by an application perspective by providing case studies in using Wikidata for projects in medicine, linguistics, or geography. However, this research seems to be scattered over different research fields in disciplines and it is challenging to develop a mental map of the existing state of the art of research. Motivated by this observation,we conducted our study, which summarize and reflects on the insights of existing research and give an overall overview of what studies have been carried out so far, and what topics needs to be explored in future research. 

A systematic mapping study provides a "map" of a research area. It helps to shape research directions by revealing existing topics which aid to identify white spots ~\cite{felderer_guidelines_2018}\footnote{A mapping study differs from a systematic literature review insofar that the later tackles a specific research question~\cite{petersen_systematic_2008}, therefore, a mapping study can be seen as a pre-study of a systematic literature review.}. It is a way of getting an overall overview of the research performed in an area of interest and classify the relevant research to get a better understanding of which areas have been covered so far and provide a baselines to assist new research efforts \cite{kitchenham_using_2011}.  
%Such a distribution provides an overview of a field to help research identified topics that are well-studied and topics that are in need of additional study'' \cite{felderer_guidelines_2018}
In our mapping study, we summarize what have been researched so far about Wikidata, when, from which origins and where they were published. We also identify which aspects of Wikidata has got more attention in the research community and which aspects are not yet given much efforts to study, by classifying and categorizing existing research from October 2012 to June 2018. Based on the search results from academic search engines, i.e. ACM, Springer Link, DBLP and Google Scholar, we identified 1,497 search results. All papers were screened and when needed, read in more detail for a more accurate decision for inclusion of the papers in the final data set. Finally, all needed information was extracted from the final set of 67 papers to answer the research questions as listed in the following section. With this mapping study, we make the following contributions: (1) we provide an overall overview of the current state of Wikidata research, (2) we identify the research areas of Wikidata where research needs to be deepen, and (3) we suggest future research areas.

This article is organized as follows. In Section 2, we explain our approach and research method in more detail. Then in Section 3, we present our findings, and discuss and reflect them in Section 4. In the final Section 5, we provide a conclusion on our research.
 
\section{Research Method}
\label{sec:method}

Our research method is motivated by our goal to provide a general overview of the field by identifying the topics that are well-studied and derive the open spots in research~\cite{felderer_guidelines_2018}. Mapping studies are insofar a suitable instrument since they provide the ground and directions of the future research as well as educate the members of a community~\cite{kitchenham_educational_2010}. 

Our study adopts guidelines for systematic mapping studies which are defined by Petersen et al.~\cite{petersen_guidelines_2015}. In the following, we describe every step to ensure that our results are comprehensible. Next, we introduce our research questions which frame the main goals of the study and inform the data collection process.

\subsection{Research Questions}
\label{subsec:rqs}

In this study we want to provide an overview of Wikidata from a research perspective. 
The peer production system Wikipedia, for example, has already drawn research from a myriad of disciplines~\cite{okoli_peoples_2012} and the question is, whether we have the same situation in the context of  Wikidata. Our research is guided by the following questions:

\begin{itemize}
    \item [\emph{RQ1}] What high-quality research has been conducted with Wikidata as a major topic or data source?
    \item [\emph{RQ2}] What types of research have been published, when (year) and where (journals or conferences)?
    \item [\emph{RQ3}] What are the origins of the research (which countries, and institutions)?
    \item [\emph{RQ4}] Which aspects of Wikidata are covered by considered research and which aspects are still to be studied?
\end{itemize}

In the following, we describe in more detail, how and where we searched articles, which papers we included or excluded respectively, and finally what categories we derived from the articles.

\subsection{Search Process and Data Sources}
\label{subsec:search}

Data collection is a crucial step in any research since findings are the direct result of the gathered data. We defined the needed keywords which is a first step for searching literature. As the noun ``Wikidata'' is only used as the name of the structured data source so far, and has no further meanings, the search string was simply selected as ``Wikidata'' in order to identify a broad range of related literature. Similarly, as Wikidata was launched in October 2012, the time range was defined from 10/2012 till 06/2018 (some search engines which did not support ``month+year'' format). The search strategy for this study is an automated search using digital libraries. We obtained Wikidata research from the ACM Digital Library (ACM DL)\footnote{ACM DL is available at: \url{https://dl.acm.org/}.}, the Springer Link Digital Library (Springer Link)\footnote{Springer Link is available at: \url{https://link.springer.com/}.}, and the Digital Bibliography \& Library Project (DBLP)\footnote{DBPL is available at: \url{https://dblp.uni-trier.de}.}. ACM DL and DBLP are bibliography search engines specifically for Computer Science. Although, Springer Link provides results from a broader range of fields such as, social sciences and humanities, we decided to extend the scope of the search in order to achieve a more holistic image of the current state of research on Wikidata from different disciplines. Thus, we included search results from Google Scholar Search Engine (Google Scholar) as well\footnote{Semantic Scholar (\url{https://www.semanticscholar.org}) is another source for Wikidata research papers, however, the filtering mechanism of this system was functioning unexpectedly and the results were not reproducible. Although we contacted the SemanticScholar team, the issue could not be solved, and thus, this search engine was not included in the study.}. 

\begin{table}[t]
\caption{\label{tab:search-results}Search results from academic search engines.}
\centering

\begin{tabular}{|l|p{1.2cm}|p{1.4cm}|p{1.3cm}|} \hline

Search Engines      & Search Results & First Screening & After Selection \\\hline
ACM 			    & 53	 		 & 53 		   	   & 	21	\\
DBLP 			    & 68	 		 & 44 		   	   & 	14	\\
SpringerLink  	    & 379	 		 & 329  		   & 	21	\\ 
Google Scholar      & 997	 		 & 699  		   & 	11  \\\hline
\textbf{Total}      & 1,497	 		 & 1,125 		   & 	67  \\\hline

\end{tabular}
\end{table}

The ACM DL searches keywords everywhere in the text, and only annual date settings are possible. We received 53 articles. Springer Link was also searched with the same keyword and time range as ACM and returned 379 results. The search interface on DBLP does not provide a time range selection, however, it returned the results from 2012 till now, which resulted in 68 papers. Google Scholar Search Engine was searched through \emph{Harzing’s Publish or Perish}\footnote{Harzing's "Publish or Perish" provides an interface to use Google Scholar and export all results in a number of formats. In this study we used the CSV format. The software is available from \url{http://www.harzing.com/pop.htm}.} software with the same criteria. The number of search results from Google Scholar was 997. This large number is caused by the fact that Google Scholar returns technical reports, white papers and theses as well. The total number of articles in the first stage was 1,497 (cp. Table~\ref{tab:search-results}).  

\subsection{Criteria Exclusion and Inclusion}
\label{subsec:inc-exc}

We defined inclusion criteria to find the most relevant research papers. The defined criteria for exclusion are duplicates, results in languages other than English and results which are not published in journals or conference proceedings, such as, websites, reports and data sets, theses and books.

\begin{figure} \includegraphics[width=\columnwidth]{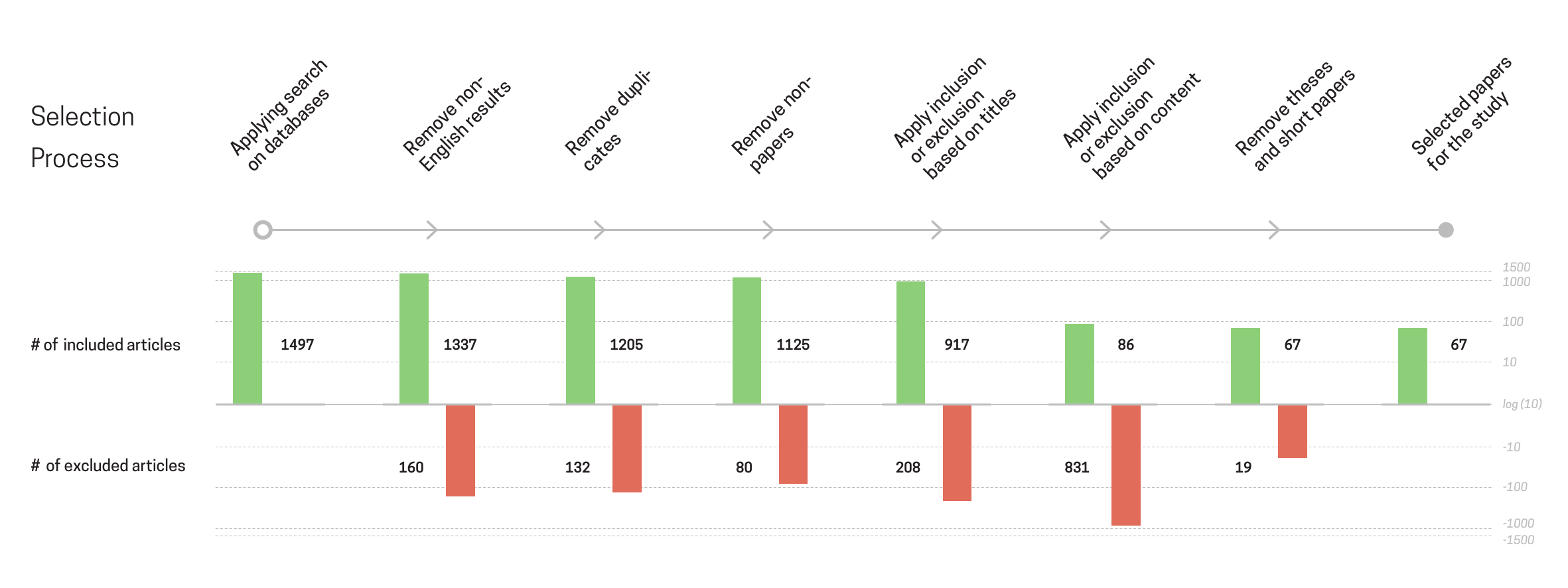}
\caption{\label{fig:screening}Article selection process and the number of included search results.}
\end{figure}

In a first step, we already excluded 160 non-English search results (145 from Google Scholar and 15 from Springer Link), second, 132 duplicates (results which were received by more than one search engine), and third, 80 non-papers (citations, refworks, reports, datasets and books). Thus, the remaining 1,125 search results were subject to an inclusion process (cp. Figure~\ref{fig:screening}).

In a second step, we included research papers only, if they are published in academic journals or conference proceedings and are full research papers with at least five pages. The latter criterion is based on the reasoning that articles with four or less pages are considered as short papers, and usually are posters, position or demonstration papers.
After applying the aforementioned criteria on the remaining 1,125 articles, 208 articles were excluded by reading the titles\footnote{The article excluded here were not caught automatically and detected after reading, which were either books, tutorials or teaching material on semantic web technologies, welcome notes of conference or workshop proceedings, blog posts or studies on Wikipedia, DBpedia or YAGO for instance. The dataset can be shared on request.} and, another 833 papers were excluded after reading the abstracts, because they were not focused on Wikidata. After reading the articles in more detail, another 17 could be identified as (bachelor and master) theses and short papers.

In total, a majority of the 1,497 found articles were excluded and only 67 papers remained in our sample. The reason for exclusion of this large number of search results were that we carried out a full text search of the term Wikidata. The search engines returned results which contained this term, even if it was used only once. As we intended to include only papers which focus solely on Wikidata, we had to exclude a large number of results. Another reason was that Google Scholar returned results which were not only papers. 

Our further discussion is based on these 67 articles. The resulting data set is available on Zotero\footnote{All papers are available in Zotero: \url{https://www.zotero.org/groups/2212336/wikidata\_study}.}. All papers are also listed in the references section. The ones marked with asterisk, are references that are not part of the mapping study.

\subsection{Data extraction}
\label{subsec:data-ext} 

Within the data extraction part of our study, we specified what data we want to extract from our data set. Having a uniform data extraction form reduces both, bias and internal validity threats. We developed a data extraction form, to answer the research questions of this study as stated in Section~\ref{subsec:rqs}.We extracted title, author(s), abstract, date, publisher to answer RQ1, RQ2 and RQ4. However, RQ3 required manual extraction of the institutions and countries where the first author of the paper had performed the research. We focused on institutions rather than the first authors themselves, because some authors published by different institutions. One author, for example, published a research paper from institution A, and later joined institution B and published there. It would be difficult to select one institution as the origin of that author. RQ4 required more insights about each research, and therefore, we read the article in more detail by focusing on the findings. The tools used for data extraction and analysis are Zotero and Microsoft Excel.
\begin{comment}
% Table 2- Data extraction form
\begin{table}
\caption{\label{tab:data-ext}Data extraction form}
\centering

\begin{tabular}{|p{5cm}|r|r|r|} \hline

Data Item  													& RQ relevance  \\\hline
Title 	 													& RQ1	 			\\\hline
Author(s) 	 												& RQ2	 			\\\hline
Abstract  	     											& RQ1, RQ4	 		\\\hline
Date		    											& RQ2	 		 	\\\hline
Publisher  													& RQ2  		     	\\\hline
Origins of the research (countries and institutions)  		& RQ3  		     	\\\hline

\end{tabular}
\end{table}
\end{comment}

\subsection{Research Paper Classification}
\label{subsec:classification}

At this stage, we read the abstract, introduction and conclusion parts of all articles to get more insights about each research for categorization. In a number of cases, further sections of the papers had to be read to get a better understanding of the scope and the topic of the paper. After analyzing 67 papers, we manually categorized the papers as shown in Table~\ref{tab:classification}. In order to define the categories, we started with descriptive labels for each paper. After reading a few papers, we tried to identify categories at a higher level of abstraction. We compared our categories throughout the reading process to make sure that our coding scheme stays consistent.

%Our categorization has three levels. We first labeled the papers discussing similar topics, then we grouped the labels, at the end, we did the categorization of groups
%based on the focus of papers on Wikidata as a KB or KG. The three main categories are: a) KB, where research papers focus on Wikidata as a structured source of knowledge, b) KG, where the graph format of Wikidata is focused and discussed, and c) KB+KG, contains the research which focus on a topic (e.g, data quality) and it would make redundant categories if the papers were further categorized as KB and KG.

\begin{table}[t]
\caption{\label{tab:classification}Classification of Wikidata research papers.}
\centering
%p{0.5cm}, p{3cm}, p{5cm}
\begin{tabular}{|p{2.6cm}|p{3cm}|p{0.7cm}|p{0.5cm}|} \hline

Sub-categories	& Labels 	& Papers 	& Sum 					\\\hline

\multirow{3}{2cm}{Community-oriented Research} &   Design Decision &	 3  	&  \multirow{3}{*}{14}	\\\cline{2-3}	 &  WD Community   &	 5   	& 	\\\cline{2-3}		
                                       
 & Multilingualism	 &	 6  &  \\\hline     

\multirow{2}{3cm}{Engineering-oriented Research}			 
& Enhancement Features  &    4 		& \multirow{2}{*}{9}  \\\cline{2-3}
& Vandalism Detection 	&	 5		& 	\\\hline 
 
\multirow{3}{2cm}{Application Use Cases} 	& Medical \& Biological Data 	&    4 		& \multirow{3}{*}{7}   \\\cline{2-3}
    					 	  				& Linguistics  			    	&    3 	    & \\\hline
\multirow{2}{2cm}{Knowledge Graph Oriented Research} & Comparison of KGs &    7 	& \multirow{3}{*}{15}   \\\cline{2-3}

 & Common issues of KGs  			  &    3   &  \\\cline{2-3} 
 &   Wikidata as Linked Data Provider    &	 5  &   \\\hline
 
\multirow{2}{2cm}{Data-oriented Research}

& Data Quality Issues 			&	9  	& \multirow{2}{*}{22}  \\\cline{2-3}
& Tools \& Datasets	         	&	13		&  \\\hline

\end{tabular}
\end{table}

%****************************************************************************************************************
\section{Overview of Data set}
\label{subsec:answeringRQs}

In this section, we describe the resulting dataset of articles in more details. First, we look at the frequency of publications, second, we determine the publication venues and third, we identify where the research was published. Finally, we look at the geographical origin of the Wikidata research.

\begin{figure} \centering \includegraphics[height=0.15\textwidth]{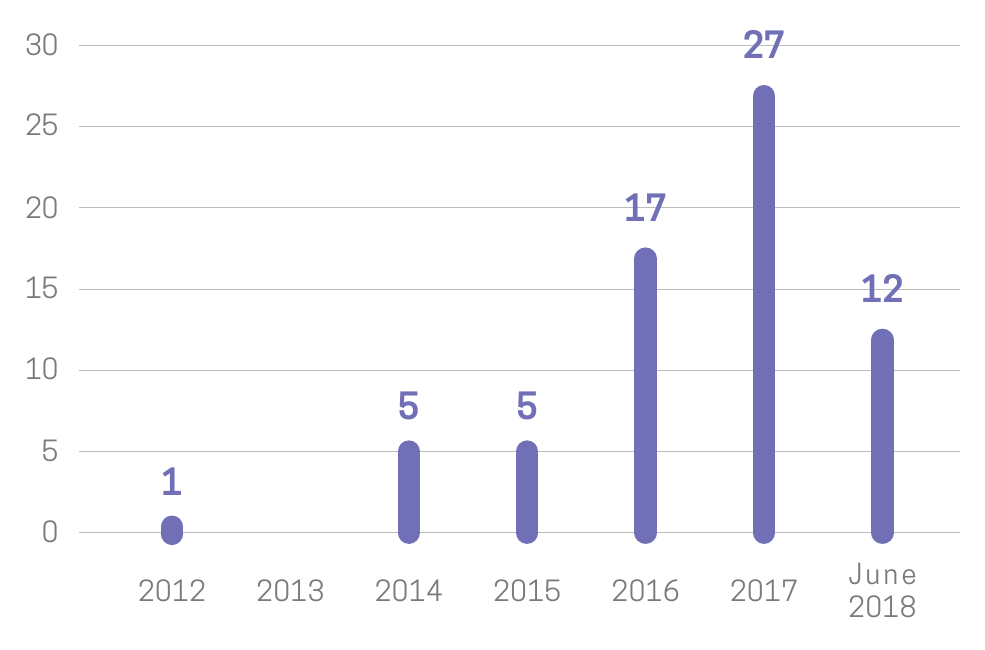}
\caption{\label{fig:frequency}Frequency of publications per year.}
\end{figure}

\subsection{Frequency of Publication}
\label{subsec:frequency}

The majority (39, 58\%) of the included 67 research papers, are recent researches from 01/2017 till 06/2018 (cp. Figure~\ref{fig:frequency}). This is an indication that Wikidata has gained more awareness in the research community. Starting from 2012, except for 2013, this number has expanded each year. Considering this growth and the number of studies until June 2018 (12), the number of research articles on Wikidata are expected to reach between 40-50 by the end of 2018.

\subsection{Publishers and Publication Types}
\label{sub:publishers}

The most popular publishers for Wikidata research are Springer with 21 articles and ACM with 19 articles, and the most popular journal for publishing Wikidata research articles is The Semantic Web Journal.
Among the 67 papers of this study, most of them (53, 79\%) are published as conference papers and the resthttps://www.overleaf.com/project/5c3f2935235d8259ff21db4e are journal articles (14, 21\%). Thus, conference proceedings are the most popular publication type in Wikidata.

The most popular conferences where Wikidata research was presented are the TheWebConf (The Web Conference)\footnote{Formarly known as International Conference on World Wide Web (WWW)}, ISWC (International Semantic Web Conference), OpenSym (The International Symposium on Open Collaboration), ESWC (Extended Semantic Web Conference)\footnote{Formarly known as European Semantic Web Symposium (ESWS)}, WSDM (ACM International Conference on Web Search and Data Mining) and MTSR (Research Conference on Metadata and Semantics Research). 

\begin{comment}

%Figure .. Most popular conferences on Wikidata research
\begin{figure} \centering \includegraphics[%
  width=0.2\textwidth]{Conferences}
\caption{\label{fig:conf}Most popular conferences on Wikidata research.}
\end{figure}
\end{comment}

\subsection{Geographical Origins of Research}
\label{subsubsec:origins-RQ3}

\begin{figure} \includegraphics[%
  width=0.5\textwidth]{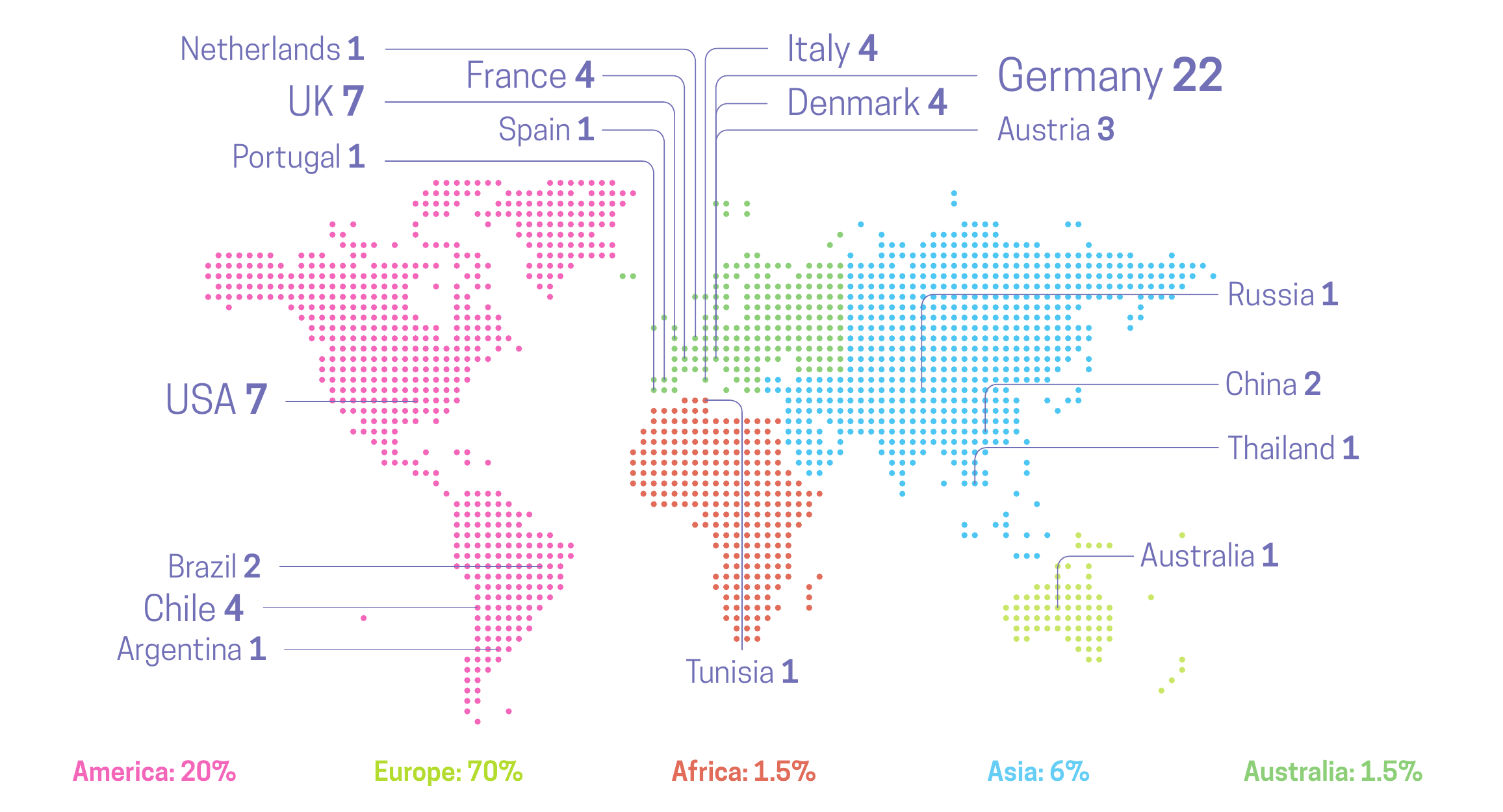}
\caption{\label{fig:map}Research contributions from countries and continents.}
\end{figure}

We found that Europe (70\%) is the dominating contributor in Wikidata research, with Germany being the leading country and United Kingdom the second. America (20\%) has also contributed in research focusing on Wikidata, with the US having the most contributions (cp. Figure \ref{fig:map}).
% Figure 5: Research contributions from countries and continents

Regarding the most active contributions of institutions, the findings show that University of Southampton has had the most contributions (7 articles), following by the Chile University (4 articles). University of Lyon, TU Dresden and TU Denmark have the same level of contributions (3 articles) on the third place, while the other contributions come across German Universities mainly.

\subsection{Research Topics of Wikidata}
\label{subsubsec:topics-RQ4}

The main research topics of Wikidata which are obtained after classification, are listed in Table \ref{tab:classification} and explained in Section \ref{sec:findings}. While, there exist a number of research topics in Wikidata, there is still potential, as we show in Section~\ref{sec:discussion}, for usage of Wikidata for a variety of purposes.

%************************************************************************************************
\section{Findings}
\label{sec:findings}

In this section, we describe the state of the art in each of the defined categories. 
The first section (cp. Section~\ref{subsec:desc}) comprises articles that look at Wikidata from a community perspective, the second section (cp. Section~\ref{subsec:newf}) contains articles from engineering perspective, the third section (cp. Section~\ref{subsec:app}) focuses on usage of Wikidata in certain fields, the fourth section (cp. Section~\ref{subsubsec:generalKGs}) discusses Wikidata from a KG perspective, and final section (cp. Section~\ref{subsec:data}) consists research on the data perspective.

%**********************************
\subsection{Community-oriented Research}
\label{subsec:desc}

Is Wikidata just another peer production system? The research in this category reflects on Wikidata's goals and features, existing design decisions (esp. multilingualism), analyzes the Wikidata community and their participation patterns.

\subsubsection{Design Decision}
\label{par:WD-intStr}
The research articles in this section provide mainly an overview on Wikidata and introduce its features and design principles.

One of the first articles on Wikidata is by Vrandečić, who motivates the need for integrating existing structured data from the various Wikipedia language versions into one single repository in order to overcome existing data inconsistencies~\cite{vrandecic_rise_2013}. The main distinguishing features of Wikidata according to Vrandečić et al. are, being available internationally and support for multilingualism, storing links to facts as a secondary database, and the ability to store contradictory facts to represent knowledge diversity~\cite{vrandecic_wikidata:_2014}.
Voß~\cite{vos_classification_2016} discusses extraction and classification of knowledge organization systems based on Wikidata.

\subsubsection{Participation Patterns of the Community}
\label{subsec:participationPatterns}

This section reflects the efforts made to understand who are Wikidata's contributors and what participation patterns do they follow. Steiner~\cite{steiner_bots_2014}, for example, develops an application which is capable of monitoring real-time edit activity of all language versions of Wikipedia and Wikidata. 

Müller-Birn et al.~\cite{muller-birn_peer-production_2015} analyze the contribution patterns of the Wikidata community to better understand whether Wikidata community participation pattern follows a peer-production approach like Wikipedia, or a collaborative ontology engineering approach. The study also describes the characteristics of the Wikidata community as, registered users, anonymous (not registered or logged in) and bots. Based on the results of this study, Cuong et al.~\cite{cuong_applicability_2016} study the dynamics of Wikidata community participation process, to know how the participation patterns of the community change over time. Piscopo et al.~\cite{piscopo_wikidatians_2017} extends this line of research by studying the participation patterns of Wikidata community members, from being an editor to becoming a community member and investigate on how these patterns evolve.
In another study Piscopo et al.~\cite{piscopo_what_2017} analyzed the relationship between group composition of bots, and humans (registered or anonymous) and the item quality in Wikidata. In their research, they focussed on the knowledge base but highlighted the importance of considering the knowledge graph in future research.

\subsubsection{Multilingualism}
\label{subsubsec:multiling}

Multilingualism is one of the design principles of Wikidata. Wikidata stores data in a language independent form and aims to provide data to anyone, anywhere in the world. This section comprises studies that focus primarily on this design principle. 

Samuel~\cite{samuel_collaborative_2017} describes the multilingual collaborative ontology development process in Wikidata by explaining the development process of a new property and its major steps from being proposed to get approved by the community and finally translated to other languages. Kaffee et al.~\cite{kaffee_glimpse_2017} study the languages covered by Wikidata. Their results suggest that most of the labels and descriptions on Wikidata are only available in a small number of languages like, English, Dutch, French, German, Spanish, Italian, and Russian. This stands in contrast to the majority of languages which have close to no coverage. Kaffee et al. in another studies, \cite{kaffee_learning_2018,kaffee_mind_2018}, investigate the generation of open domain Wikipedia summaries from Wikidata in ``underserved languages'' to overcome uneven content distribution. Ta et al. \cite{ta_model_2014} propose a mechanism to enrich Wikidata multilingual content by retrieving ``semantic relations based on alignment between info-box properties and Wikidata properties in various languages''.
Sáez et al.~\cite{saez_automatically_2018} investigate the development of ``fully automatic methods'' where info-boxes for Wikipedia can be generated from Wikidata descriptions.

%**********************************
\subsection{Engineering-oriented Research}
\label{subsec:newf}

This section contains all articles that suggest approaches and features that enhance Wikidata's functionality. These features are programmed for two main purposes: first, for improving the quality by adding new data or by interlinking with other sources, and second, for vandalism detection.

\subsubsection{Enhancement Features}
\label{subsubsec:enhancement}

Wikidata's functionality has evolved gradually with the needs of the community. This section contains research that proposes approaches that ease the process of adding data to Wikidata either manually, or by using external data sources.

Zangerle et al.~\cite{zangerle_empirical_2016} evaluate recommender algorithms, which assist Wikidata contributors in the process of data insertion through property recommendation. Pellissier Tanon et al.~\cite{pellissier_tanon_freebase_2016} introduces the Primary Sources Tool\footnote{For more information, please check~\url{https://github.com/google/primarysources}.} to facilitate the migration of the content from Freebase to Wikidata. 
Sergieh et al.~\cite{mousselly_sergieh_enriching_2016} propose an approach to bridge the missing linguistic information gap of Wikidata by aligning Wikidata with FrameNet\footnote{FrameNet is a lexical database of the English language. For more information, please check ~\url{https://framenet.icsi.berkeley.edu/}.} lexicon. Hachey et al.~\cite{hachey_learning_2017} present a neural network model for mapping structured and unstructured data and investigate the generation of Wikipedia biographic summary sentences from Wikidata. 

\subsubsection{Vandalism Detection}
\label{subsubsec:vandalism}

Wikidata provides data for Wikipedia and other Wikimedia projects; thus, the integrity and correctness of data is of high importance. Vandalism detection is, therefore, an essential aspect of a knowledge repository and directly influences the data quality and trustworthiness of a KB. In the following, we provide an overview of research that focuses on detecting vandalism and other efforts for making Wikidata more robust.

In their study, Heindorf et al. \cite{heindorf_vandalism_2016} present a new machine learning-based approach for the automatic detection of vandalism in Wikidata.
Sarbadani et al. \cite{sarabadani_building_2017} develop a vandalism detection mechanism for Wikidata by adapting methods from the Wikipedia vandalism detection literature and extending it to Wikidata’s structured knowledge base. The mechanism used identifies damaging changes and classifies edits as vandalism in real time, using a machine classification strategy. 

The ACM International Conference on Web Search and Data Mining, held the competition for developing vandalism detection mechanisms for Wikidata, the WSDM Cup 2017.\footnote{More more information, please check ~\url{https://www.wsdm-cup-2017.org/}.}  The main goal of this competition was to develop a model for detecting malicious or similarly damaging edits. As a result of participation in WSDM Cup 2017 both, Crescenzi et al.~\cite{crescenzi_production_2017} and Grigorev et al.~\cite{grigorev_large-scale_2017}, presented their own vandalism detection mechanisms\footnote{The competition received five submissions: 1) Buffaloberry by Crescenzi et al.~\cite{crescenzi_production_2017}, 2) Conkerberry by Grigorev~\cite{grigorev_large-scale_2017}, 3) Loganberry by Zhu et al.~\cite{zhu_wikidata_2017}, 4) Honeyberry by Yamazaki et al.~\cite{yamazaki_ensemble_2017}, and 5) Riberry by Yu et al.~\cite{yu_2017}. We included two of the submissions only, because~\cite{zhu_wikidata_2017} and~\cite{yamazaki_ensemble_2017} are short papers and~\cite{yu_2017}  is not published.}. Crescenzi et al.~\cite{crescenzi_production_2017} reflected on previous work of Heindorf et al.~\cite{heindorf_vandalism_2016}, an automatic data mining approach for vandalism detection in Wikidata. Grigorev et al.~\cite{grigorev_large-scale_2017} present an approach based on a linear classification model, which according to authors, is faster compared to other existing approaches.

The evaluation of the proposed vandalism detection approaches at the WSDM Cup 2017, is done in~\cite{heindorf_overview_2017}. Heindorf et al.~\cite{heindorf_overview_2017} evaluate their four baseline approaches\footnote{The four baseline approaches are: 1) Wikidata Vandalism Detector (WDVD) approach from~\cite{heindorf_vandalism_2016}, 2) FILTER, a second baseline which contains trained data from 01.05.2013 to 30.04.2016, 3) ORES, the re-implementation of the approach in~\cite{sarabadani_building_2017}, and 4) META, a combination of all approaches in~\cite{heindorf_wsdm_2017}.} along the five submissions. The study finds that the best approach is a semi-automatic scenarios ``where newly arriving revisions are ranked for manual review'' is from~\cite{crescenzi_production_2017}, while, the best approach in a fully automatic detection scenario ``where the decision whether or not to revert a given revision is left with the classifier'' is the baseline approach by the Wikidata Vandalism Detector (WDVD) system~\cite{heindorf_vandalism_2016}.

%********************************************************************************************

%\subsection{Analytic research}
%\label{subsubsec:analytical}

%The research papers in this category analyze Wikidata internal issues, and provide solutions to enable and enhance Wikidata to be used in accordance to semantic web technologies.

%%%%%%%%%%%%%%%%%%%%%%%%
\subsection{Application Use Cases}
\label{subsec:app}

From the beginning, Wikidata received many attentions from members of various research fields. Many articles described possible use cases for utilizing Wikidata as a central data hub, as we see in the next section.

\subsubsection{Medical and Biological Data}
\label{subsubsec:med}

Recently, especially medical and biological projects have started using Wikidata as a backend data source, to facilitate data exchange, mapping, and consumption. Mitraka et al.~\cite{mitraka_wikidata:_2015}, for example, propose the usage of Wikidata for addressing the crucial challenges in disseminating and integrating knowledge in life sciences contexts, by linking genes, drugs and diseases. 
Pfundner et al.~\cite{pfundner_utilizing_2015} have specified an automated process to integrate data from ONC’s\footnote{The Office of the National Coordinator (abbrev. ONC) for Health Information Technology is a division of United States' Department of Health and Human Services.} high priority DDI\footnote{DDI stands for Drug-Drug Interaction, i.e. the effect change of one drug on body by another drug.} list into Wikidata. The authors aim to integrate the data from ONC into Wikidata and then use Wikidata to display the integrated data in articles of different Wikipedia language versions. Burgataller-Muehlbacher et al.~\cite{burgstaller-muehlbacher_wikidata_2016} import all human and mouse genes, and all human and mouse proteins into Wikidata to improve the state of biological data, and facilitate data management and data dissemination using the WDQS of Wikidata.

Although, Wikidata is greatly being used in bioinformatics, it is still a challenging task for biologists to use it efficiently. One major issue is for example, that the ``structured query languages like SPARQL are not commonly part of a researcher’s toolkit''. Thus, E. Putman et al.~\cite{e._putman_wikigenomes:_2017} describe WikiGenomes, a web application based on Wikidata, that facilitates the ``consumption and curation of genomic data by the entire biomedical researcher community''. WikiGenomes provides access to the centralized biomedical data and a simple user interface for non-developer biologists.

\subsubsection{Linguistics}
\label{subsubsec:dict}

Wikidata is also used in the linguistics field, either as a dictionary, or proposing further approaches for linking lexical datasets or relation extraction.

Turki et al.~\cite{turki_using_2017} propose to adopt Wikidata as a dictionary which can be used across multiple dialects of the Arabic language. The authors emphasize that the Arabic language has many dialects and these dialects are not all mutually intelligible, and each one of them has its morphological and phonological and even semantic and lexical particularities. The study explains how it is possible to convert Wikidata into a multilingual multidialectal dictionary for Arabic dialects and describes how Wikidata (as a multilingual multidialectal dictionary for Arabic dialects) can be used by computational linguistics and computer scientists in the Natural Language Processing of the varieties of the Arabic language. Nielsen et al. \cite{nielsen_linking_2018} describe an ongoing effort for linking ImageNet\footnote{``ImageNet is an image dataset organized according to the WordNet hierarchy'' (\url{http://image-net.org/}.} WordNet\footnote{WordNet is a large lexical database of English and contains and groups nouns, verbs, adjectives and adverbs in the form of sets of cognitive synonyms (synsets). For more information \url{https://wordnet.princeton.edu}.} synsets to Wikidata.
Yu et al.~\cite{yu_meronymy_2017} present a new approach for meronym relations extraction in Wikidata, which is, building a 13-dimensional feature vector for each hyperlink to be classified with different classification algorithms, based on all 13 different three-node motifs. The high interest of this community might have one driver for the development of the Wikibase Lexeme extension which allows for modeling lexical entities. From 2018, Wikidata includes this new type of data: words, phrases, and sentences.

\subsection{Knowledge Graph Oriented Research}
\label{subsubsec:generalKGs}

Wikidata is maintained by an active community of contributors who create a large amount of structured data. The knowledge base relies on the MediaWiki infrastructure. At the meantime, Wikidata's structured data is stored in RDF and is accessible through SPARQL. Wikidata belongs, therefore, to a group of other general purpose knowledge graphs, such as DBpedia, YAGO, and Cyc.

\subsubsection{Wikidata as Linked Data Provider}
\label{subsec:WD-semweb}

We summarize all articles that propose approaches for storing Wikidata's structured data in RDF and on the other hand, suggest how projects in Wikimedia's ecosystem can use the RDF data.

Erxleben et al.~\cite{erxleben_introducing_2014} argue that despite being the data platform in the Wikimedia ecosystem, Wikidata provides its data not in RDF, which affects Wikidata's popularity in the Semantic Web community negatively. Thus, the authors propose an RDF encoding for Wikidata and introduce a tool~\footnote{For more information, please check~\url{https://www.mediawiki.org/wiki/Wikidata_Toolkit}.} for creating such RDF file exports. Similarly, Hernández et al.~\cite{hernandez_reifying_2015} compare various options for reifying RDF triples from Wikidata, and building on that study the efficiency of various database engines for querying Wikidata~\cite{hernandez_querying_2016}.

From 2014, Wikidata stored its data in RDF based on the Erxleben et al.-mapping~\cite{erxleben_introducing_2014} and provided the data via an SPARQL endpoint, the Wikidata Query Service (WDQS)\footnote{The WDQS is available here~\url{https://query.wikidata.org/}.}. Bielefeldt et al.~\cite{bielefeldt_practical_2018} analyzed the access logs from SPARQL endpoint and separate the bot-based from human-based traffic. As expected, the human part is smaller and shows clear trends, e.g. correlated to time of day, in comparison to the bot-based part which is ``highly volatile and seems unpredictable even on larger time scales''.

Yang et al.~\cite{yang_relation_2017} uses the data for improving Wikipedia. They discuss that KGs can help machines to analyze plain texts, and propose a Relation Linking System for Wikidata (RLSW) which links the Wikidata KG to data in plain text format in Wikipedia.

\subsubsection{Comparison of KGs}
\label{comparisonKGs}

Next, we discuss articles which compare Wikidata with other general domain knowledge graphs. Ringler \& Paulheim~\cite{ringler_one_2017}, for example, study DBpedia, Freebase, OpenCyc, Wikidata and YAGO knowledge graphs to find similarities and differences of these KGs. Färber et al. compare in their research, KGs from a data quality perspective (~\cite{farber_comparative_2015, farber_linked_2016}). Razniewski et al.~\cite{razniewski_but_2016} discuss the challenges of asserting completeness in KGs, and outline possible solutions. The authors propose a framework for finding the most suitable KG for a given setting. Abian et al.~\cite{abian_wikidata_2018} compare Wikidata and DBpedia structured data sources, based on the criteria defined in the main data quality frameworks. In a similar study, Thakkar et al.~\cite{thakkar_are_2016} compare DBpedia and Wikidata from a quality assurance perspective and have found that based on the majority of relevant metrics, the quality of Wikidata is higher than DBpedia.
Data quality of Wikidata has also been studied from a KG perspective, as in study from Gad-elrab et al.~\cite{gad-elrab_exception-enriched_2016} which discuss that KGs like DBpedia, Freebase, YAGO and Wikidata are inevitably incomplete. To address this, the authors analyze the former approach of data correlations and propose a method to overcome the problems with former approach.

\subsubsection{Common Issues of KGs}
\label{commonKGs}

Ismayilov et al.~\cite{ismayilov_wikidata_2015} describe the integration of Wikidata into the DBpedia Data Stack in order to use Wikidata through DBpedia extractors. In their study, Chekol \& Stuckenschmidt~\cite{chekol_towards_2018} discuss that KGs, such as YAGO, Wikidata, NELL, and DBpedia, already contain temporal data (facts together with their validity time). The authors propose a ``bitemporal'' model for knowledge graphs, to record the data extraction time from other sources. Currently, only NELL records this time, while, Wikidata only contains the time which is valid about a fact. In another study, Krötzsch discusses the modern knowledge representation technologies and their advantages in information management, such as description logics, and their contribution to knowledge graphs, and motivates Wikidata as a use case~\cite{krotzsch_ontologies_2017}.

\subsection{Data-oriented Research}
\label{subsec:data}

This section contains research which make use of data from the Wikidata knowledge base and the knowledge graph. Some papers belong to KB and some to KG, while all focus on their defined category. We organized the papers in two categories: (1) data quality quality aspects of Wikidata, and (2) the development of new tools and datasets.

\subsubsection{Data Quality}
\label{subsubsec:quality}

The research highlighted in this section, is concerned with improving the data quality of the knowledge graph by providing tools for Wikidata's knowledge base.

Prasojo et al.~\cite{prasojo_managing_2016} discuss “COOL-WD”, a tool for supporting the completeness lifecycle of Wikidata and allow to produce and consume completeness data by ``data completion tracking, completeness analytics, and query completeness assessment.'' Augenstein~\cite{augenstein_joint_2014} discusses that KBs are far from complete and proposes information extraction methods to populate missing knowledge from Web pages to KBs. Galárraga et al.~\cite{galarraga_predicting_2017} 
investigate ``different signals to identify the areas where the knowledge base is complete'' and experiment in Wikidata and YAGO to generate completeness information automatically.
Razniewski et al.~\cite{razniewski_doctoral_2017} introduce the problems and limitations of properties in Wikidata and propose entity-specific property ranking for Wikidata.
Ahmeti et al.~\cite{ahmeti_assessing_2014} and Balaraman et al.~\cite{balaraman_recoin:_2018} propose and develop Recoin, a relative completeness tool for evaluating completeness of entities in Wikidata. Recoin uses information from the class structure of the knowledge graph, in order to recommend possible properties for an item on the Wikidata user interface.   
 Brasileiro et al. \cite{brasileiro_applying_2016} discuss the quality of taxonomic hierarchies in Wikidata to have a consistent data model and representation schema.
Piscopo et al. (\cite{piscopo_provenance_2017, piscopo_what_2017_1}) analyze Wikidata quality from the provenance perspective, the relevance and authoritativeness of Wikidata external references.

\subsubsection{Tools \& Datasets}
\label{subsubsec:ToolsnDS}

This category contains research that resulted in the development of new tools, which mainly use Wikidata as a backend data source. Ontodia~\cite{wohlgenannt_using_2017}, for example, is a ''simple and free online OWL and RDF diagramming tool``. Scholia~\cite{nielsen_scholia_2017} is a tool for handling scientific bibliographic information through Wikidata, and NECKaR~\cite{geis_neckar:_2017} is a named entities classifier based on Wikidata, which provide also a Wikidata-based named entity data set.
Ferrada et al.~\cite{ferrada_querying_2018} present a new web interface for IMGpedia dataset which can query more than 6 million images of IMGpedia through Wikidata, while, Diefenbach et al. (\cite{diefenbach_wdaqua-core0:_2017, diefenbach_wdaqua-core1:_2018}), present and discuss WDAqua-core which is a new Questions Answering component which uses DBpedia and Wikidata. Veen et al.~\cite{veen_linking_2016} use Wikidata to improve access to the collection of Dutch historical newspapers. 

More recently, some effort has been investigated to synchronize the data beween OpenStreetMap and Wikidata. Leyh et al. \cite{leyh_interlinking_2017} discuss the opportunities and challenges of Wikidata as a central integration facility by interlinking it with OpenStreetMap. Almeida et al.~\cite{almeida_where_2016} introduce a tool that harmonizes street names from OpenStreetMap\footnote{For mire information please check: \url{https://www.openstreetmap.org/}} and the entities they refer to in Wikidata. Another study, Thornton et al.~\cite{thornton_modeling_2017} explore the potential of Wikidata to serve as a technical metadata repository and how it provides distinct advantages for usage in the domain of digital preservation. 

There are also datasets which were developed based on Wikidata for different purposes. Nielsen et al.~\cite{nielsen_inferring_2018} construct a dataset containing pairs of digital photos of objects for a multi-modal knowledge representation.
Klein et al.~\cite{klein_monitoring_2016} develop ``Wikidata Human Gender Indicators'' (WHGI), a biographic dataset, to monitor gender related issues. Spitz et al.~\cite{spitz_so_2016} present an approach for constructing a network of locations from Wikipedia by computing the similarity of locations based on their distances and linking it to Wikidata as a knowledge source.

\section{Discussion}
\label{sec:discussion}

Our mapping study shows an increase in the number of published research articles per year, which indicates the growing interest of the research community on Wikidata (cp. Section~\ref{subsec:frequency}). The articles have a prevalence of computer science articles which we expected from the chosen databases which are mainly Computer Science related (ACM, Springer Link, DBLP). However, by including Google Scholar, we expected to identify more research from disciplines such as sociology or communication science. Unfortunately, our results suggest that this approach was less successful. However, as other peer production communities Wikidata provides a valuable opportunity to deepen our understanding of existing community practice. It might be interesting, for example, to explore existing difference to Wikipedia. Furthermore, within various Wikipedia language versions, there is still a resistance to use Wikidata. Further research is needed, to better understand existing reservations. Another interesting less studied aspect in Wikidata is the existing human-bot-collaboration~\cite{muller-birn_peer-production_2015}. Wikidata might be, besides Wikipedia, an interesting use case to better understand the social-technical infrastructure of a peer production community. 

Our results suggest that research on Wikidata seems to be entirely concentrated on specific institutions, such as the University of Southampton or the Universidad de Chile, or countries, for example, Germany and USA (cp. Section~\ref{subsubsec:origins-RQ3}). It might be the origin of Wikidata as a European project initiated by members of the Semantic Web community which causes that research on Wikidata is more popular in Europe. We wonder, how this western perspective on knowledge representation might exclude other understandings of knowledge. For example, the indigenous peoples give their knowledge orally from generation to generation. Research, which deals with the question of how this knowledge or the potential occurrence of such knowledge can be represented, would undoubtedly be useful to achieve the aim for becoming a global universal knowledge base, which can be used by anyone for any purpose~\cite{vrandecic_wikidata:_2014}. 

While there have been studies on the multilingualism aspect of Wikidata, the data is still not present in every language. Current findings show that there are some dominant languages (e.g. English, French, German, Spanish), while, many other languages as `underserved' (cp. Section \ref{subsubsec:multiling}). This indicates that, although, there have been some efforts in addressing the issue of uneven languages distribution, further studies are needed to overcome the language gap in Wikidata. Furthermore, these studies focus on the descriptions and labels of an item. It might be interesting to understand better when Wikidata's data model fails because a one-to-one relationship between two words from different languages is not possible.

Continuous evolution is one of the design decisions of Wikidata, which means Wikidata grows with its community and tasks, and new features are deployed incrementally~\cite{vrandecic_wikidata:_2014}. The findings suggest only little research on improving the usability of the user interface.
User studies concerning aspects such as the learnability or explainability are still rare on Wikidata. From the authors own experiences on conducting Wikidata workshops, it can be said, that people struggle with understanding Wikidata's central concepts, for example, the difference between a class and an instance. It seems that Wikidata has still untapped potential in becoming accessible for non-technical experts. 

Many efforts are made to sustain and improve the quality and completeness of data in Wikidata (cp. Section~\ref{subsubsec:quality}). One issue in this context is, for example, the handling of vandalism and data integrity. 
In the context of data quality, we call for more research on the effects of plurality, i.e., the co-existence of contradictory information, in order to enhance the trustworthiness of Wikidata content. However, if anyone can add contradictory information, further research is needed to provide such mechanisms in the user interface as well in the WDQS for providing this information in a possible format. 

As opposed to Wikidata, Wikipedia is studied from a variety of disciplines, such as, humanities (e.g, history, literature, philosophy), logic and mathematics, natural sciences (biology, chemistry), social sciences (e.g. communications, education, economics, law, journalism) and interdisciplinary (anthropology, computer science, health, industrial ecology and information science) [14]. While, Wikidata has the competence to be used in different disciplines, the investigations are needed to find out whether Wikidata can be beneficial in the same areas where Wikipedia was used.
Even though our study reveals the usage of Wikidata in various contexts, the uses cases come from the biomedical domain and linguistics mainly (cp. Section~\ref{subsec:app}). It might be valuable to see more use cases from other disciplines, such as social sciences or humanities. It might be valuable, for example, to use Wikidata in educational or museum settings.  

\section{Limitations of Research}
\label{subsec:validityThreat}
%As discussed in \cite{petersen_guidelines_2015}, considering possible validity threats are a quality criterion for a mapping study. 
In this section, we discuss a number of validity threats this study is subject to. To overcome descriptive validity (research design) threats adequate information, we provided all from data collection to data analysis as detailed as possible in the constrains of the publication format. Further, the research was designed in way to minimize the number of missing literature by including results from Google Scholar. Although, Google Scholar results contained many irrelevant results (e.g. citations), it was meaningful nevertheless, because we captured 11 articles not found in the other libraries. Theoretical validity is achieved by capturing the most relevant literature and controlling bias in the data extraction and classification steps~\cite{petersen_guidelines_2015}. We addressed theoretical validity by including peer-reviewed articles only, and to reduce bias, the results were carefully checked by the second author. Interpretive validity is achieved when the conclusions are the result of the given data~\cite{petersen_guidelines_2015}. One of the threats in drawing conclusion is researcher bias, which has been controlled by the second author review. Repeatability can be achieved by providing detailed information of each research step and the data. We provide all our data, the search log on github and the final article sample on Zenodo.\footnote{\textit{This information will be included in the final version of this paper.}}

\section{Conclusion}
\label{sec:conclusion}

In this mapping study, we have provided an overview of existing research about Wikidata. We identified existing research topics in this field and described potential new research topics for future studies. The literature was collected from digital libraries and academic search engines, and the selected papers were categorized based on research focus relevance.
Research publications accelerated every year which is an indication of the interest of research community. Most of the research contributions come from Europe so far, thus, Wikidata is still predominantly used and studied from a Western perspective. This affects, for example, multilingualism and knowledge diversity in Wikidata. Although, data in Wikidata is available in various languages simultaneously, this applies only for a selected number of languages, i.e. many languages have very few or no coverage in Wikidata. Thus, future directions of research on Wikidata could be to: a) focus on multilingualism aspect of Wikidata and overcome language gaps, b) study knowledge diversity and the effect of plurality on data trustworthiness of Wikidata, c) research on improvement of the usability of user interface, and d) investigate the usage of Wikidata in various disciplines and study it from non-technical perspective.

% The acknowledgments section is defined using the "acks" environment (and NOT an unnumbered section). This ensures
% the proper identification of the section in the article metadata, and the consistent spelling of the heading.
\begin{acks}
This part will be provided in the camera-ready version.
\end{acks}

%
% The next two lines define the bibliography style to be used, and the bibliography file.
\bibliographystyle{ACM-Reference-Format}
\bibliography{sample-sigconf}

% 
% If your work has an appendix, this is the place to put it.
%\appendix

%Appendix comes here...


\end{document}